\documentclass[aps,groupedaddress,superscriptaddress,preprint,nofootinbib,pre]{revtex4-1}

\usepackage{graphicx}
\usepackage{color}
\usepackage{amsmath}
\usepackage{paralist}
\usepackage{dcolumn}
\usepackage{booktabs}
\usepackage{amsfonts}
\usepackage{amssymb}
\usepackage[acronym]{glossaries}
\usepackage{xcolor}

\usepackage{xr-hyper}

\newacronym{WAN}{WAN}{World Airline Network}

\usepackage[final]{pdfpages}

\begin{document}

\title{Emergence of core-peripheries in networks}

\author{T. Verma} 
\affiliation{ ETH
  Z\"urich, Computational Physics for Engineering Materials, Institute
  for Building Materials, Wolfgang-Pauli-Strasse 27, HIT, CH-8093 Z\"urich,
  Switzerland}

\author{F. Russmann} \affiliation{ ETH
  Z\"urich, Computational Physics for Engineering Materials, Institute
  for Building Materials, Wolfgang-Pauli-Strasse 27, HIT, CH-8093 Z\"urich,
  Switzerland}

\author{N. A. M. Ara\'ujo \footnote{Correspondence and requests for materials should be addressed to N. A. M. A. (nmaraujo@fc.ul.pt)}} \affiliation{Departamento de F\'isica, Faculdade de Ci\^encias, Universidade de Lisboa, P-1749-016 Lisboa, Portugal, and Centro de F\'isica Te\'orica e Computacional, Universidade de Lisboa, 1749-016 Lisboa, Portugal}

\author{J. Nagler} \affiliation{ ETH
  Z\"urich, Computational Physics for Engineering Materials, Institute
  for Building Materials, Wolfgang-Pauli-Strasse 27, HIT, CH-8093 Z\"urich,
  Switzerland}
  
\author{H. J. Herrmann} \affiliation{ ETH
  Z\"urich, Computational Physics for Engineering Materials, Institute
  for Building Materials, Wolfgang-Pauli-Strasse 27, HIT, CH-8093 Z\"urich,
  Switzerland} \affiliation{Departamento de F\'isica, Universidade
  Federal do Cear\'a, Campus do Pici, 60455-760 Fortaleza, Cear\'a,
  Brazil}

%\date{\today}

\maketitle

% \begin{abstract}

A number of important transport networks, such as the airline and trade networks of the world, exhibit a characteristic core-periphery structure, wherein a few nodes are highly interconnected and the rest of the network frays into a tree. Mechanisms underlying the emergence of core-peripheries, however, remain elusive. Here, we demonstrate that a simple pruning process based on removal of underutilized links and redistribution of loads can lead to the emergence of core-peripheries. Links are assumed  beneficial if they either carry a sufficiently large load or are essential for global connectivity. This incentivized redistribution process is controlled by a single parameter which balances connectivity and profit. The obtained networks exhibit a highly resilient and connected core with a frayed periphery. The balanced network shows a higher resilience than the World Airline Network or the World Trade Network, revealing a pathway towards robust structural features through pruning.

% \end{abstract}
\pagebreak
% \pacs{}

\section*{Introduction}
In today's world we want to fly everywhere. Despite higher fuel prices and a wider consciousness for reducing carbon emissions, airplane travel is on the rise globally and is predicted to grow even further in the future \cite{IATA}. Events like the shutdown of the entire European airspace, due to the eruption of the Icelandic volcano, Eyjafjallaj\"okull, 
have demonstrated the importance of %depicted how important it is to take note of the 
efficiency and reliability of the airline traffic \cite{Budd} and other transport networks, be it trade, Internet or trains. 
 
An ideal point-to-point network topology would ensure the fastest transfer of loads in a transport network. However, the real world imposes costs on transport networks and their actual structure is a result of a complex interplay of (among other factors) economic considerations of involved parties as well as political ties between different regions. For instance, most major airlines, nowadays, employ a hub-and-spoke philosophy in which passengers are routed through a few central airports, depending on the size of the airline's fleet. In recent years, however, especially low-cost airlines (for example, Ryanair in Europe) have rediscovered the point-to-point philosophy, providing non-stop flights wherever sufficient demand exists \cite{Zanin}. This results in a denser and more clustered network as opposed to a hub-and-spoke one. 

One of the remarkable features of the \gls{WAN} is its small core (consisting of about $2.5\%$ of the airports) that is almost fully connected and surrounded by a vast periphery that is nearly tree-like and connected to the core through many regional and national hubs \cite{Verma}. This block arrangement is prominently known as the core-periphery structure \cite{Borgatti,Holme,Silva,Csermely} which was also reported for other infrastructure networks, such as the World Trade Network \cite{Fagiolo,Rossa}, the autonomous Internet network \cite{Rossa} and the financial interbank lending markets \cite{Elliott}, where %with 
 the fraction %number 
 of peripheral nodes %in the range
 varies from $45\%$ to $85\%$. % of the network. 
  Rombach {\it et al.} \cite{Rombach} have also found similar structures for friendship, voting and collaboration networks and Avin {\it et al.} for other social networks \cite{Avin}.

The reason behind core-peripheries is still unclear. Some transport network \cite{Black,Gastner} models have been based on a greedy optimization
of a particular evaluation function of distance, cost or time. None of
the above studies, however, could reproduce the core-periphery
structure. 

We hypothesize that the core-peripheries are a result of a
naturally existing state of the dynamics of networks that are driven by
a balance between functional connectivity and load-based profit. As an
illustration of this hypothesis, commercial airlines will very likely
cancel a direct link if the number of passengers does not compensate for
the associated costs. Here, we start with a Utopian network where each
node is connected to every other node. Underutilized links are pruned
and the load of such links is redistributed to guarantee the load
transfer between nodes. Through this pruning model, we demonstrate that core-periphery structures can be obtained.

\section*{Results}
\subsection*{Model}
Generally, in transport networks, load is anything that needs to be transported from one place to another. We start with an ideal fully-connected and undirected network where load pertaining to a pair of nodes can be transferred bidirectionally (a full description of the algorithm is given in Supplementary Methods \cite{Dijkstra}). 

We represent the network using an adjacency matrix $A_{ij}(N,V)$ with $N$ nodes and $V$ links representing whether or not there exists a direct link between any pair of nodes. Our reference network contains $N = 1000$ nodes. Since we are interested in transport networks, we consider that a link is characterized by its load $l_{ij}$, cost $c_{ij}$, and physical length $d_{ij}$ (Euclidean distance between nodes, in km, taken randomly from a Gaussian distribution, $\mu = 8.369 \times 10^{3}; \sigma = 4.954 \times 10^{3}$. The nodes are spread around a sphere of the size of the Earth - see Supplementary Figure 1 and Supplementary Note 1). 

We define the profit of a link connecting nodes $i$ and $j$ as
\begin{equation}\label{eq:profit}
u_{ij} = b_{ij} - c_{ij},
\end{equation}
where $b_{ij}$ is the benefit arising from a link and $c_{ij}$ is the cost of establishing and maintaining the said link. Since the load of a link is a proxy for the benefit it accrues, we set $b_{ij} = l_{ij}$. For simplicity, we assign the same cost to every link with a dispersion to accommodate for heterogeneity in the network; $\vartheta \equiv c_{ij}$ and $c_{ij}=(1+\delta_{ij})c$, where $\delta_{ij}$ is a
uniformly distributed random number in the range $\lbrack-a;a\rbrack$. In
particular, we consider the cases $a=\{0, 0.05, 0.1\}$. We obtain good quantitative agreement for the three cases, showing that our results are robust to heterogeneity in
the parameter $c_{ij}$ (see Supplementary Figure 2 and Supplementary Note 2). Varying $\vartheta$ from the minimum load, we systematically prune links of negative profit, starting with the least loaded ones. An underutilized link that is necessary for maintaining global connectivity is not removed and classified as essential.

Once a link is pruned, its load is redistributed through the next best (shortest-path) alternative, which potentially turns these alternative links more beneficial than they were before. In the case where several paths are of the same length, one is chosen at random. The load redistribution process can be explained in two steps.
Firstly, when a link is removed, the load is routed through the next
shortest path available between the nodes. Secondly, every link on the
next available path will have to absorb the incoming load as it moves
from source to sink. The reason for choosing the shortest path as the
next available path is because normally in a transport network, the
length of travel times and in most passenger driven networks,
convenience is of primary importance to both the consumers and service
providers. However, a robustness analysis of two other alternatives
(random path, second shortest-path) shows that
core-periphery features are observed in the critical window and the
robustness of the networks in different regimes remains the same (see
Supplementary Figure 3). The pruning process eventually gives rise to a network only comprising essential links.

To distribute the loads, we introduce an observable called the
popularity, $p_i$, for each node $i$, characterizing its
importance for the network. The popularity of a node is initially
randomly chosen from a uniform distribution in the range $\left[1/3,1
\right]$, and alternatively from a scale-free distribution, $P(p) \sim
k^{-\gamma}$, to contrast and compare the effect of initial conditions
on our model (see Supplementary Figure 4 and Supplementary Note 3).  Subsequently, the
initial load on any link is defined as the product of popularities of
the nodes involved (see Supplementary Figure 5 and Supplementary Note 4 for an in depth understanding of this relationship),
\begin{equation}\label{eq:load}
l_{ij} = p_i p_j \ \ .
\end{equation} 
The popularity of each link remains intact with the pruning process. 
However, the load of each link dynamically changes as the load of
removed links is redistributed. We have examined several other load
functions (such as $l_{ij} = p_i + p_j$, $\log(p_i + p_j)$, $\log(p_i
p_j)$ and $\exp(p_i p_j)$) and found no significant dependence of the
main findings on the load function. In addition, we have also used a
specific and more conventional case of load, betweenness centrality (see Supplementary Figure 6 and Supplementary Note 5). As will be evident in the Results section, the existence of core-peripheries
remains the same. However, the load and its redistribution are needed
(and critical) to find the core-periphery structure.

\begin{figure*}[t]
\includegraphics[width=\columnwidth]{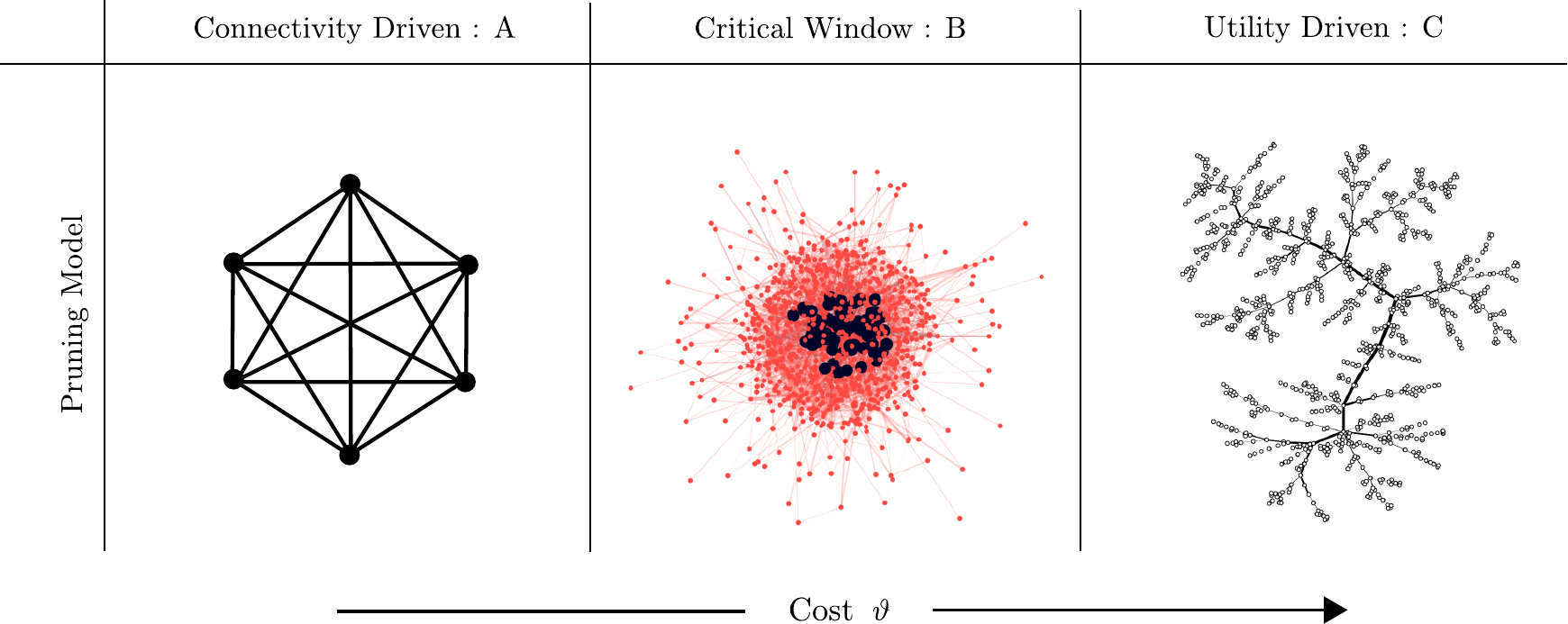}
\caption{
{\bf 
Schematic of the network classes obtained by our algorithm.
} For vanishing cost, the network is fully connected (network A of six nodes - shown for simplicity) resembling the initial network. For significantly high cost, the network is tree-like, exhibiting no loops (network C of $10^3$ nodes). In between, the proposed pruning process generates a network (network C of $10^3$ nodes) with a core-periphery structure. The network in regime B was obtained for cost, $\vartheta = 0.92$, corresponding  to a peak in the core-periphery measure (details in the text). For the central network, the layout was generated by applying the Fruchterman-Reingold algorithm \cite{Fruchterman}. Colors show the difference in magnitude of coreness with black indicating the core and red, the periphery.}\label{fig:visuals}
\end{figure*}
We run the above algorithm and analyze the structure using standard network techniques. The pruning process coupled with the load redistribution mechanism gives rise to three distinct families of network structures (see fig.\ \ref{fig:visuals}), one of which strongly resembles the features of a core-periphery structure.

To identify and analyze the core-periphery structure we use the t-core decomposition, as proposed in Ref. \cite{Verma}. Similar to the k-core decomposition \cite{Dorogovtsev1}, this method progressively prunes a network by recursively removing nodes that are part of the least number of triangles. The decomposition assigns the removed nodes a ``coreness'', $t$, and places them in different shells, $t = 0,1,2\ldots$, where a shell, $t$, has nodes that are part of at least $t$ triangles. Since triangles enhance the resilience of load transfer and this method recovers subgraphs at every shell that are more and more densely connected, the method uncovers a hierarchical ordering. More specifically, the load passing between a pair of nodes in a transport network can be redistributed with only one change in case the direct link becomes unavailable. A node that is part of the fully-connected core of a network will be able to transfer its load through many alternatives (as many as there are nodes in the core) to accommodate for a faulty link. Thus, the $t$-core measure is especially suitable to assess which nodes belong to the core or the periphery. 

To compare networks of different sizes, we define the relative coreness %, $\tau$, by
\begin{equation}
\tau = \frac{t}{T},
\end{equation}
where $T = \frac{(N-1)(N-2)}{2}$ is the maximum possible coreness of a node in a network of $N$ nodes.

To perform a more aggregate level analysis where core-periphery structure across different networks can be studied, we  focus on the core-periphery (CP) measure, a dimensionless quantity defined as 
\begin{equation}\label{eq:lambda}
\lambda = (\tau_{\max} - \tau_{\min})\frac{S_{\tau_{\min}}}{S_{\tau_{\max}}},
\end{equation}
where $\tau_{\max}$ and $\tau_{\min}$ stand for the maximum and minimum relative coreness found in the network, respectively, and $S_{\tau_{\min}}$ and $S_{\tau_{\max}}$ for the number of nodes that were assigned the respective coreness. 
A network with a genuine core-periphery structure will have both, many nodes with low coreness (periphery) and a few nodes with high coreness (core). For example, the empirical \gls{WAN} has a ratio, $\frac{S_{\tau_{\min}}}{S_{\tau_{\max}}} = 42.5$, that is much larger than unity, suggesting 
%thereby indicating
 the presence of very few nodes in the core, compared to the periphery. Thus, a high ratio %description of this 
% factor in $\lambda$ 
 indicates a particularly pronounced core-periphery, and a low value, the lack of a core-periphery.
 The rationale behind definition (\ref{eq:lambda}) is based on qualitative experience with the empirical WAN, which distinctly maximizes $\lambda$ as there are very few nodes in the core and the majority of nodes fall in the periphery. Moreover, the difference in the relative coreness between core and periphery ($\tau_{\max} - \tau_{\min}$) is large.

\subsection{Regimes}
The cost, $\vartheta$, is varied as an independent tunable parameter and the properties of the model networks are investigated as a function of this parameter. Specifically, we systematically increase the value of the cost, starting from the minimum load and until only essential links remain, namely links necessary to keep global connectivity.

%It is clear 
%from 
Our pruning process, depending on the value of $\vartheta$, %it is evident
 necessarily leads to %exhibit %occurs %will be 
a crossover between different regimes of networks. Say that $l_{\min}$ and $l_{\max}$ are the least and most loaded links in the initial network, respectively. The regimes are:
\begin{enumerate}
\item Connectivity Driven (A) $\vartheta \leq l_{\min}$ - In this case, no links fall below the cost and hence no pruning takes place. It is apparent that this regime will essentially have only a fully-connected network (the reference network we begin with). Networks in this regime maximize connectivity but their profit is diminished (eq. \ref{eq:profit}). 
\item Core-Periphery (B) $l_{\min} < \vartheta \leq l_{\max}$ - In this regime, the network undergoes the most rapid changes in its structure. All the links that fall below the cost are removed sequentially and the load is redistributed to the remaining network. Nodes gain more traffic and the links that get pruned give rise to a variable core-periphery character. This character is not always prominent in the entire regime and depends strictly on the value of $\vartheta$. An example is shown in fig.\ \ref{fig:visuals}(B).
\item Profit Driven (C) $l_{\max} < \vartheta$ - This regime shows extreme structural changes in the network. Most links get pruned except the ones essential for connectivity - eventually giving rise to a tree-like structure towards the end of this regime, illustrated in fig.\ \ref{fig:visuals}(C). Since we attach the same cost to each link, the cost of the network scales monotonically with the number of links. Thus, networks in this regime have the minimum possible cost. 
\end{enumerate}

\begin{figure*}[ht]
\includegraphics[width = 0.9\columnwidth]{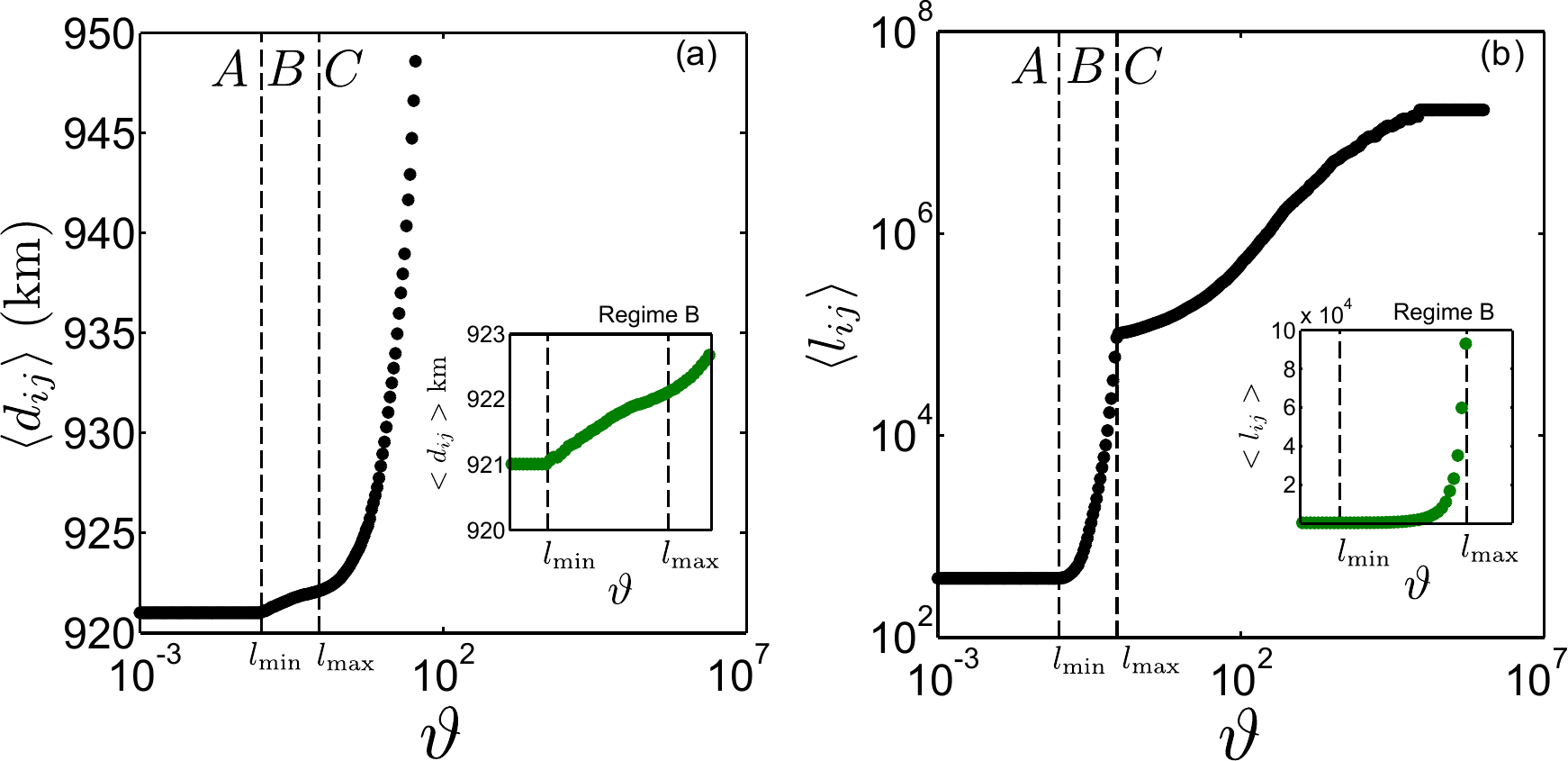}
\caption{
{\bf Average shortest path $\langle d_{ij} \rangle$ in km and average load $\langle l_{ij} \rangle$ dependence on cost $\vartheta$.}
We observe three different regimes as a function of the cost.
In (a) the average shortest path length remains relatively stable while the load (a proxy for benefit) as shown in (b)  increases drastically in regime B. 
The insets of both figures are blow-ups of regime B.
In (a) a slight increase in the shortest path in regime B is observed while in (b) the benefit increases by a large magnitude pointing to the inevitable compromise between connectivity and profit. 
Data are averages over 100 realizations.
\label{fig:kink}}
\end{figure*}
Upon removing links and redistributing their loads onto the remaining links, the modularity \cite{Leicht,Newman3}, average shortest path length \cite{Watts1} and average load per link increase; see fig.\ \ref{fig:kink}, while the average degree and average clustering coefficient decrease \cite{Watts1}. This indicates that communities start emerging while keeping beneficial links intact and sacrificing the ones that lead to a shorter path for transfer of loads. The clustering in the network decays due to a periphery that is slowly emerging. It is worth noting that these curves exhibit a kink at $\vartheta = l_{\max}$ (see fig.\ \ref{fig:kink}, Supplementary Note 6 and Supplementary Figures 7--9 for other characteristic properties). With a small increase in the average shortest path (fig.\ \ref{fig:kink}a), the average load on the remaining links increases (fig.\ \ref{fig:kink}b), thereby making them more significant for the network. At $\vartheta = l_{\min}$ the network changes rapidly and links start getting pruned as they fall short of justifying their existence. Around $\vartheta = l_{\max}$, we observe that the network exhausts its pruning capabilities. The links that are removed now are the most loaded and hence transfer much more load to other links thereby slowing down the pruning process considerably. 

Additionally, in regime A, since no link is pruned, the average shortest path length remains constant. As the pruning process becomes effective, the average shortest path slightly increases with the cost (regime B). By contrast, in regime C, the average shortest path increases exponentially with the cost. Note that as illustrated in the
Supplementary Figures 10 and 11, the fraction of essential links required to
ensure global connectivity is small unless the costs are very high,
indicating that the constraint of global connectivity does not affect
the network's proclivity towards core-peripheries.

\subsection{Core Size}
\begin{figure*}[ht]
\includegraphics[width = 0.9\columnwidth]{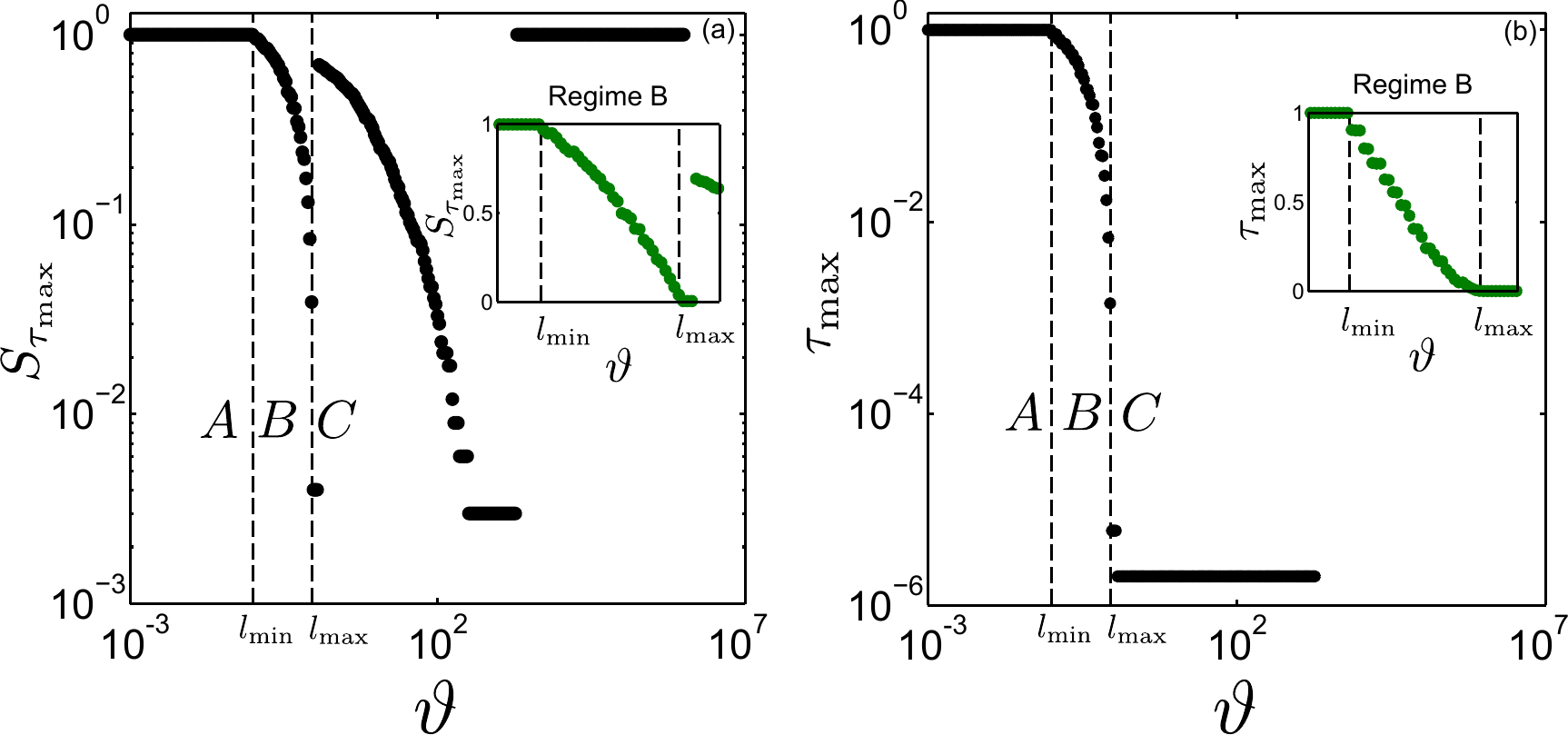}
\caption{
{\bf Characteristic metrics of %the 
t-core decomposition.}
Core size, $S_{\tau_{\max}}$, and relative coreness, $\tau_{\max}$, vs the cost, $\vartheta$. 
(a) shows a decay in the size of the core in regime B for increasing cost. 
Core size increases again abruptly in the transition between regimes B and C as the pruning mechanism slows down. 
(b) illuminates upon the comparison of the relative coreness of the core between a fully connected network in regime A and a core-periphery observed in regime B. 
The insets of both figures are blow-ups of regime B. The core of the network in regime B has a much lower coreness which decays continuously with increasing cost until the network becomes a tree. Data are averages over 100 realizations. \label{fig:coreness}}
\end{figure*}
\begin{figure}[ht]
\includegraphics[width = 0.5\columnwidth]{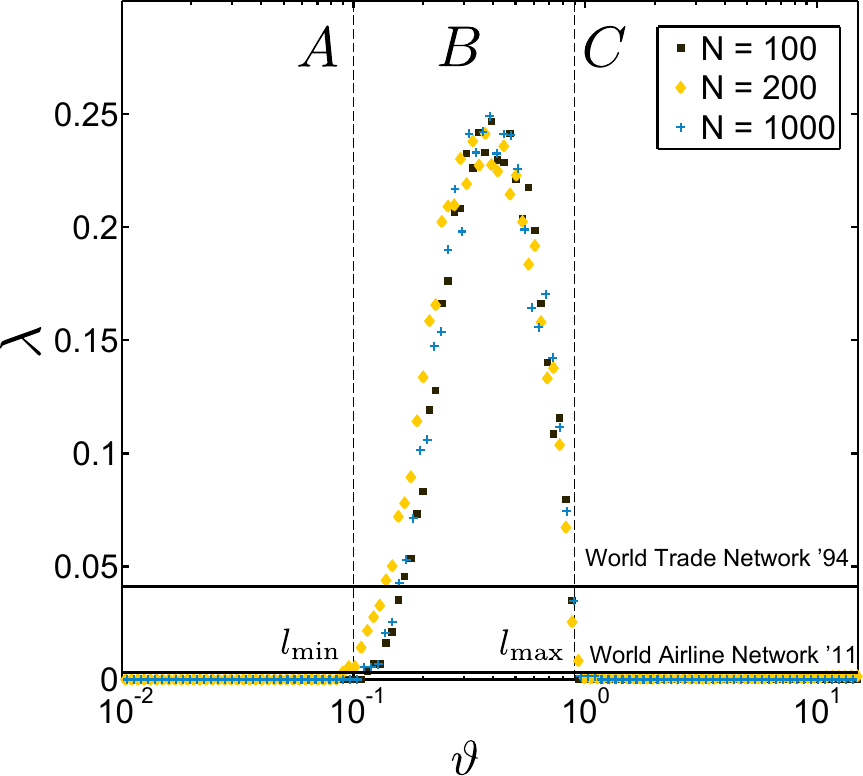}
\caption{
{\bf Core-periphery measure $\lambda$ as a function of  
 $\vartheta$ for different system sizes $N$.}
Modeled networks in regime B have a high value of $\lambda$ owing to their core-periphery characteristic and resilience. The world trade network from year $1994$ lies close to $\lambda = 0.041$ and the world airline network from the year $2011$ is at $\lambda = 0.0032$. The trade network is only comprised of $80$ nodes, whereas, the airline network has close to $3500$ nodes. 
Data are averages over 100 realizations. \label{fig:lambda}}
\end{figure}

 
A t-core decomposition was performed at every value of $\vartheta$ to assess the network's core-periphery properties. We measured the size of the core, $S_{\tau_{\max}}$, and the maximum relative coreness of the network, $\tau_{\max}$, as a function of the cost. Figure \ref{fig:coreness} shows that in regime A, where the network is still fully connected, the core consists of the entire network with a very large coreness since there are many triangles. On the other hand, in regime C, the tree-like network is sparsely connected such that it is essentially segregated into one shell at coreness, $t = 0$. Remarkably, between regimes B and C, the core size exhibits a discontinuity. The network undergoes a transition from a state where the size of the core is comparable to the system size but is of small coreness to a state with a small core and relatively large coreness. Since the empirical \gls{WAN} is known to have a small core size of approximately $2.3\%$ but high inter-connectivity within the core \cite{Verma}, it should be found in regime B with $l_{\min} < \vartheta \leq l_{\max}$, where the value of $\lambda$ is largest (see fig.\ \ref{fig:lambda}). $\lambda$ is close to zero in regime A and C, because we have a fully-connected network in A and a tree-like one in C. However, in regime B, where $\lambda \approx 0.25$ is maximum, we find a periphery emerging which is held together by the core in the middle (see fig.\ \ref{fig:visuals}(B)). In this region, the difference in the relative coreness between core and periphery ($\tau_{\max} - \tau_{\min}$) is huge and the ratio of the number of nodes in the periphery to that of the core is much larger than unity ($\frac{S_{\tau_{\min}}}{S_{\tau_{\max}}} \gg 1$). The world trade network \cite{Smith} and the \gls{WAN} \cite{Openflights} are also included in fig.\ \ref{fig:lambda} for comparison (solid horizontal lines). The trade network is only comprised of $80$ nodes, whereas, the airline network encompasses about $3500$ nodes. These networks exhibit a lower core-periphery measure, $\lambda$, since there is a high cost for building networks. In contrast, a network that has no cost (or less cost) attached can comprise many more triangles within its core, consequently depicting a higher value for $\lambda$. In order to understand the physical depth of the quantity coreness, $\lambda$, we first discuss
two limits of $\lambda$: a fully connected network (regime A) and a
tree-like structure (regime C). In both cases $\lambda=0$. We tested another null configuration starting with a fully connected network of our main model where links are removed at random until
the network turns into a tree (no more pruning is possible). As shown in
Supplementary Figure 12, by contrast to the results with load
redistribution, when links are simply removed at random, there is no
well-defined maximum for $\lambda$, thus core-periphery structures do
not emerge at any stage (Supplementary Note 7).

\subsection{Coreness Distribution}
\begin{figure}[ht]
\centering
\includegraphics[width = 0.5\columnwidth]{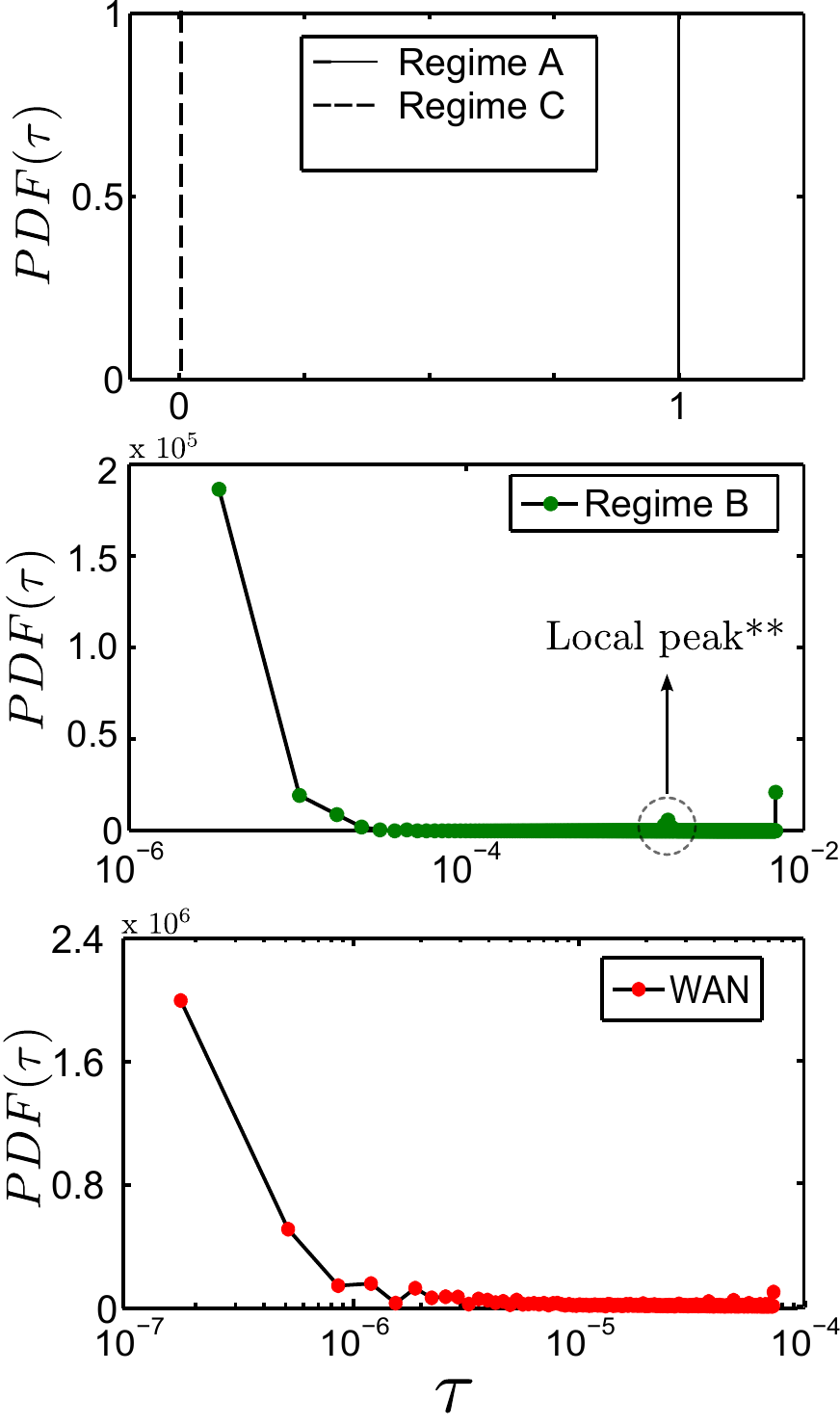}
\caption{
{\bf 
Probability density functions of coreness of different regimes and the empirical WAN.}
Regime B, for cost $\vartheta = 0.92$, that maximizes the value of core-periphery measure (independent of system size $N$), $\lambda = 0.248$ (fig.\ \ref{fig:lambda}), and the real-world network exhibit a core-periphery structure. The density functions show the probability of having a shell with relative coreness $\tau$ (relative to a fully connected network). Data are averages over 100 realizations.
\label{fig:histograms}}
\end{figure} 
To evaluate the core-periphery properties of the networks, we calculated the probability density function (PDF) of the relative coreness of some exemplary model networks in each regime as well as the empirical WAN. Figure \ref{fig:histograms} shows the PDFs of the relative coreness of networks in each regime. Qualitatively, the core-periphery structure is visible in regime B networks. The periphery consists of many nodes with small coreness; probability dropping with increasing coreness (notice the semi-logarithmic scale). 

The coreness densities of the networks from regimes A and C exhibit a markedly %completely 
different behavior. In the case of a fully connected network (regime A), it consists of a single peak at $\tau = 1$ and for the tree-like network (regime C), of a single peak at $\tau = 0$. Hence the entire network is segregated into one shell following the t-core decomposition. Due to their simplicity, the PDFs for regimes A and C are grouped in one plot. 

Figure \ref{fig:visuals} illustrates the structural difference between the core-periphery network of regime B and the tree-like network of regime C. It is immediately evident how the core nodes (in black) are highly interconnected as they are grouped closely together by the force directed Fruchterman-Reingold algorithm \cite{Fruchterman}. The algorithm uses spring-like attractive forces to attract the nodes that have a link between them, while simultaneously repulsive forces of charged particles are used to separate all pairs of nodes. This arrangement allows us to distinguish core from periphery. In the empirical \gls{WAN} network, the core is spread over continents or different regions of the world (see Supplementary Figure 13).

\subsection{Resilience}
Transportation networks in our globalized world have not resulted from
a centralized optimization procedure. Most networks have resulted from
the superimposition of many locally optimized networks and accretion of
regional networks, providing for a globalized way to travel. In such
scenarios, it is non-trivial to establish a common ground for measuring
resilience. We use a basic measure, often used in the past to
qualitatively assess the efficiency of a network \cite{Schneider} to removal of
nodes. 

\begin{figure*}[ht]
\centering
\includegraphics[width = 0.7\textwidth]{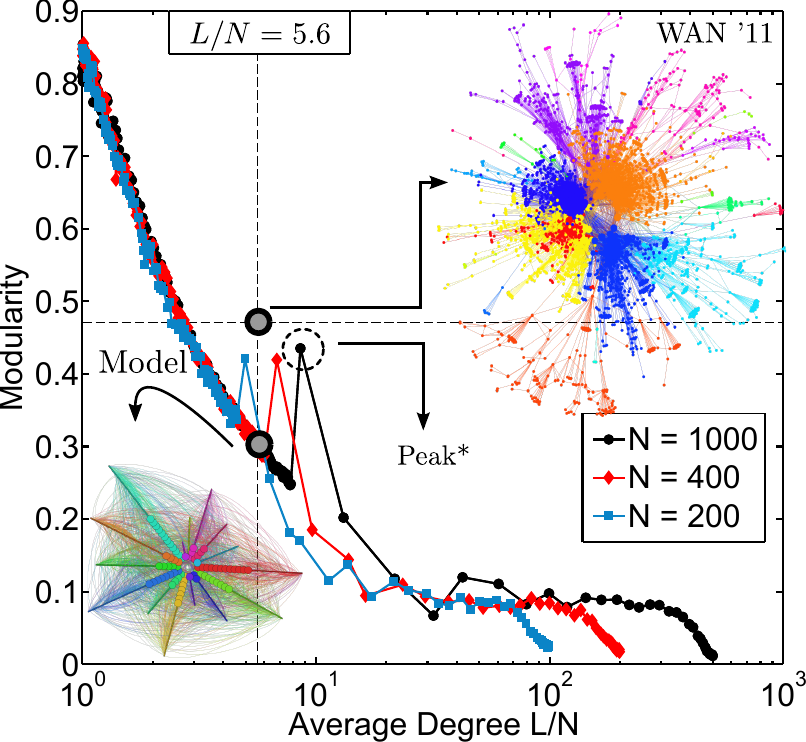}
\centering
\caption{\protect\rule{0ex}{5ex}
{\bf Modularity as a function of %the %vs
 average degree.} % of the networks.}
The model networks show a peak* in the modularity for an average degree close to the World Airline Network. This peak is due to the increase in coreness of the network as the core collapses and a larger core takes shape (see fig. 5 - local peak** observed in the distribution of coreness for modeled networks). For the same average degree, $L/N = 5.6$, the model generates many interconnected modules while the World Airline Network shows little or no links between modules. Different colors represent different communities and the size of the nodes classify them into core (large) or periphery (small). Data for system sizes $N = 200, 400, 1000$ are averages over 100 realizations.}
\label{fig:robustness}
\end{figure*}
We compare the robustness of our modeled networks - for the same average degree - with the empirical WAN. As presented in Ref. \cite{Verma}, the empirical network is very sensitive to the removal of high degree nodes and the size of the largest component drops very quickly (Supplementary Note 8 and Supplementary Figure 14). However, a model network in regime B appears more robust owing its topological strength to a strongly connected periphery where peripheral nodes have a few redundant links between each other. Figure \ref{fig:robustness} illustrates that 
the modularity of the network seems to result 
from the peripheral linkages, a topological feature that indicates the
strength of intra-community links over links across communities. This
is a grave factor contributing to its abrupt diminishing robustness. Our
model produces robust networks that accrue benefit to network elements
without compromising on the connectivity of these elements. In addition,
the modularity peaks are a result of the increase in coreness of the
network as the core collapses and a larger core takes shape (see fig. \ref{fig:histograms}
- local peak** observed in the distribution of coreness for modeled
networks). Furthermore, a detailed robustness analysis for various network
sizes shows that the change in robustness does not depend on the network
size and follows the same pattern for all network sizes (Supplementary Figure 15). 
For the same average degree, $L = N = 5/6$, the model
generates many interconnected modules while the World Airline Network
shows little or no links between modules (increment in modularity). In
other words, the model networks have lower modularity compared to the
world airline network which also has a larger average shortest path
length, giving rise to more tightly knit modules.

\section{Discussion}\label{sec:Discussion}
We have presented a model producing the qualitative nature of the core-periphery structure observed in many real world networks. Remarkably, this is possible by dynamically allowing the failed links to redistribute their loads and the network's effort to increase its profit, as two processes working on the network. We have also taken into account the costs imposed due to the spatial nature of such networks, by considering Euclidean distances between the nodes to define the new routes for the redistribution of loads. Simulating these processes on a network with no other fundamental assumptions, we obtain for a wide range of cost values, a small but densely interconnected core and a vast periphery.

Our pruning process not only produces core-periphery networks
but also reveals different network regimes. The crossover between
these regimes can be modeled using only a single cost-based parameter,
$\vartheta$. This parameter can be varied to show many interesting
properties of the modeled networks. For instance, when a core-periphery
structure is present, the average load on a link (a proxy
for the benefit of the link) increases, while the average shortest path
length between any two nodes (a proxy for convenience of load transfer)
stays stable. Additionally, connectivity is optimized in regime A 
where everything is connected and profit is optimized in regime C, 
according to the construction of our model. However, note that regime B balances 
these two real world considerations and, interestingly, we find most real-world networks to exist in this region as well.

Though, not all networks are planned, their current condition is
dictated by a variety of rules. Our efforts do not reproduce every kind
of network verbatim and do not try to fully describe the evolutionary
process of a network but give a plausible explanation for understanding
profit-driven core-periphery networks. We not only produce the
core-periphery character of networks, but also show that modeled
networks are more resilient to removal of nodes compared to the
empirical example of the world airline network. This resilience can be attributed to the less modular structure of the modeled networks. Since our modeled networks are stable and resilient to removal of nodes, it is natural to ask whether our approach could be used to design cost-efficient and resilient infrastructure networks, something policy makers might centrally control. 

The process of pruning a globally connected %of large-scale %fully connected 
network %s %with global connectivity 
fundamentally differs from the %underlies our model and not 
bottom-up growth many real networks have undergone. Schneider {\it et al.} developed a pruning model which reproduces well many topological properties of protein interaction networks \cite{Schneider1}. Inspired by this strategy of preferential depletion, our model mimics core-periphery networks closely.
Transport networks with a geographical dependence try to optimize faster connectivity with demand induced profit. 
%Citing an example from reality, 
An example includes the world airline network that is a possible outcome of individual airline networks competing and cooperating (wherever profitable) with each other. On the other hand, the networks of large carriers like Star Alliance could approximate the picture of a global network in which our model could make suggestions for improvements assuming the partners in such an alliance are able and willing to cooperate with each other. 

Lastly, Peixoto {\it et al.} \cite{Peixoto} show that the most robust topology against random failures is a core-periphery structure. By studying the
percolation properties of arbitrary large-scale networks using
robustness as the most significant force for driving the system, the
authors show that a core-periphery network is the case of maximum
entropy. Our non-equilibrium approach depicts that a network in regime
B (critical window) will be highly robust in comparison to real
networks. Louf {\it et al.} \cite{Louf} have proposed a cost-benefit driven
optimization model based on physical distances in transport networks to
study their formation. An interesting revelation of their work is that
cost driven network optimization leads to a hub-and-spoke structure,
different from a core-periphery structure in our model. Louf {\it et al.}
carried out the addition of links on a static system where the distances
dictate the future of links. Our model differs from this in a way that a
dynamic redistribution of loads is taken into account which encapsulates
the collective nonlinear effects of various local load redistributions
around the network. The interplay between load redistribution and
profit provides a plausible explanation for core-peripheries in
transport networks. We believe that our framework can be extended to
other networks that are based on profit maximization.

%% == end of paper:

%% Optional Materials and Methods Section
%% The Materials and Methods section header will be added automatically.

%% Enter any subheads and the Materials and Methods text below.

\section{Methods}

We ran tests for various system sizes, namely, $N = 100, 200, 400, \ldots 1000$ and for each system $100$ randomly selected samples were considered.

\begin{acknowledgments}

We acknowledge financial support from the ETH Risk Center, European
Research Council through Grant FlowCSS No. FP7-319968. and Portuguese Foundation for Science and Technology (FCT) under Contracts nos. EXCL/FIS-NAN/0083/2012, UID/FIS/00618/2013 and IF/00255/2013.

\end{acknowledgments}

\includepdf[pages={1-19}]{SI.pdf}

\end{document}